# THE EFFECTS OF DOUBLY IONIZED CHEMISTRY ON $SH^+$ AND $S^{+2}$ ABUNDANCES IN X-RAY DOMINATED REGIONS

N. P. Abel[1], S. R. Federman[2], and P. C. Stancil[3]

## Abstract

Recent laboratory measurements for the $S^{+2}$ + $H_2$ reaction find a total rate coefficient significantly larger than previously used in theoretical models of X-ray dominated regions (XDRs). While the branching ratio of the products is unknown, one energetically possible route leads to the $SH^+$ molecule, a known XDR diagnostic. In this work, we study the effects of $S^{+2}$ on the formation of $SH^+$ and the destruction of $S^{+2}$ in XDRs. We find the predicted $SH^+$ column density for molecular gas surrounding an Active Galactic Nucleus (AGN) increases by as much as 2 dex. As long as the branching ratio for $S^{+2}$ + $H_2$ → $SH^+$ + $H^+$ exceeds a few percent, doubly ionized chemistry will be the dominant pathway to $SH^+$, which then initiates the formation of other sulfur-bearing molecules. We also find that the high rate of $S^{+2}$ + $H_2$ efficiently destroys $S^{+2}$ once $H_2$ forms, while the $S^{+2}$ abundance remains high in the $H^0$ region. We discuss the possible consequences of $S^{+2}$ in the $H^0$ region on mid-infrared diagnostics. The enhanced $SH^+$ abundance has important implications in the study of XDRs, while our conclusions for $S^{+2}$ could potentially affect the interpretation of Spitzer and SOFIA observations.

[1]Department of Physics, University of Cincinnati, Cincinnati, OH, 45221 npabel2@gmail.com

[2]Department of Physics & Astronomy, University of Toledo steven.federman@utoledo.edu

[3]Department of Physics & Astronomy and the Center for Simulation Physics, University of Georgia stancil@physast.uga.edu

## 1 Introduction and Background

The importance of doubly ionized species in interstellar chemistry has been theorized for over thirty years. Dalgarno (1976) pointed out that doubly charged ions may react with $H_2$ to form simple molecular ions. Calculations by Langer (1978) found $CH^+$ could efficiently be formed in gas exposed to the diffuse X-ray background through this process and argued this could be the solution to the longstanding "$CH^+$ problem" in the diffuse interstellar medium (ISM). While energetically possible, laboratory measurements of $C^{+2} + H_2 \rightarrow CH^+ + H^+$ (Smith & Adams 1981) found the reaction coefficient to be negligibly small, although the possibility of CH+ formation via $C^{+2}$ is still thought possible (Sternberg, Dalgarno, & Yan 1997; Yan 1997). Recent calculations of the chemical structure of X-ray dominated regions (henceforth XDRs, Maloney, Tielens, & Hollenbach 1996) surrounding Active Galactic Nuclei (AGNs) and young stellar objects (YSOs) show significant abundances of doubly and singly charged ions such as $Ne^+$ to be co-spatial with $H_2$. This suggests that doubly charged ions could play an important role in interstellar chemistry.

Laboratory studies show some $X^{+2} + H_2$ reaction rates are fast. Chen, Gao, & Kwong (2003, henceforth CGK) found a total reaction rate coefficient for $S^{+2} + H_2$ $\rightarrow$ products of $1.58 \pm 0.13 \times 10^{-9}$ cm$^3$ s$^{-1}$ at 1077 K. Gao & Kwong (2003) found $C^{+2}$ +

H$_2$ proceeds at a rate of 8.77±0.71 ×10$^{-11}$ cm$^3$ s$^{-1}$ at 2630 K (within 15% of the value determined by Smith & Adams 1981). While neither study determined branching ratios, Yan (1997, henceforth Y97) argued (incorrectly we believe) that the charge transfer rate coefficients of S$^{+2}$ and C$^{+2}$ with H$_2$ are small. The XDR calculations of Stäuber et al. (2005) and Meijerink & Spaans (2005) only consider S$^{+2}$, C$^{+2}$ charge transfer reactions in their chemical scheme, using rate coefficients of 10$^{-15}$ and 10$^{-13}$ cm$^3$ s$^{-1}$ respectively, based on Y97. These rates are 10$^3$ – 10$^6$ times slower than the total measured reaction rate of these species with H$_2$. The more recent calculations of Meijerink, Glassgold, & Najita (2007) for X-ray irradiated protoplanetary disks do consider the CGK rate, but modeled the 6716, 6731 Å [S II] doublet ratio and not emission from [S III].

In the case of S$^{+2}$ + H$_2$, there are at least four exoergic reaction paths. These are :

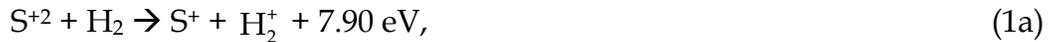
$$S^{+2} + H_2 \rightarrow S^+ + H_2^+ + 7.90 \text{ eV}, \tag{1a}$$

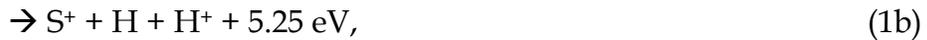
$$\rightarrow S^+ + H + H^+ + 5.25 \text{ eV}, \tag{1b}$$

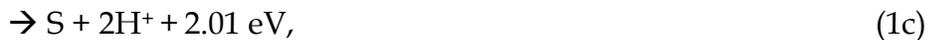
$$\rightarrow S + 2H^+ + 2.01 \text{ eV}, \tag{1c}$$

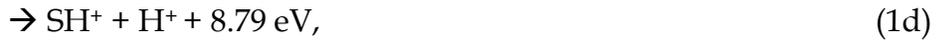
$$\rightarrow SH^+ + H^+ + 8.79 \text{ eV}, \tag{1d}$$

where the excess product energies refer to the ground state of each channel. Regardless of the branching ratios, the large measured rate coefficient for reaction (1) means H$_2$ efficiently destroys S$^{+2}$ in XDRs. However, depending on the branching ratio of this reaction, process (1) could be important in forming

$SH^+$. Y97 investigated the effects of reaction (1d) with a rate of $10^{-9}$ cm$^3$ s$^{-1}$ and found a significant increase in the $SH^+$ abundance with a decrease in the $S^{+2}$ abundance in XDRs. The possible importance of $S^{+2}$ on the formation of $SH^+$ and the [S III] mid-IR emission lines was also discussed in Sternberg, Dalgarno, & Yan (1997) and Y97.

We investigate the degree with that $S^{+2}$ reacting with $H_2$ contribute to $SH^+$ formation, and how $S^{+2}$ is destroyed in XDRs. We also study how $N(S^{+2})$ varies with changes in the branching ratio of reaction (1).

## 2 SH$^+$ Chemistry

In an XDR, $SH^+$ can be formed via several channels, including $S + H_3^+ \rightarrow SH^+ + H_2$ and $S + HCO^+ \rightarrow SH^+ + CO$ (Stäuber et al. 2005). Another reaction that leads to $SH^+$ formation is $S^+ + H_2 \rightarrow SH^+ + H$. This reaction is only efficient in warm environments, as it is endoergic with a temperature barrier of 9860 K. The endothermicity of this reaction is also likely why $SH^+$ has never been definitively detected in the diffuse ISM (Savage, Apponi, & Ziurys 2004). Observations of the 3360 Å line by Magnani & Salzer (1989, 1991) place an upper limit to $N(SH^+)$ of $(1.8 - 9.8) \times 10^{12}$ cm$^{-2}$. James Clerk Maxwell Telescope observations of YSOs were also restricted to upper limits due to blending by $SO_2$ lines (Stäuber et al. 2007).

If we ignore $S^+ + H_2 \rightarrow SH^+ + H$, then we can write down an expression for the total $SH^+$ formation rate in gas exposed to X-rays, which is:

$$3.3\times10^{-10}n(S)[7.9n(H_3^+)+n(HCO^+)]+k_1n(S^{+2})n(H_2) \qquad (2)$$

In (2), we used rate coefficients taken from the latest version of the UMIST database (Woodall et al. 2007) where $k_1$ is the rate coefficient for (1). Since $n(H_2)$ is typically much larger than $[7.9n(H_3^+)+n(HCO^+)]$, SH$^+$ formation via S$^{+2}$ + H$_2$ will be important as long as $3.3\times10^{-10}n(S)$ is not several dex greater than $k_1n(S^{+2})$.

Destruction of SH$^+$ is largely due to dissociative recombination, SH$^+$ + e → S + H. The rate coefficient for this reaction is $2\times10^{-7}\times(T/300)^{-0.5}$ cm$^3$ s$^{-1}$. SH$^+$ is also destroyed through UV dissociation, with a rate of $3\times10^{-10}\times G_0\times\exp(-1.8A_V)$ s$^{-1}$, where $G_0$ is the strength of the Far-Ultraviolet (FUV) field between 6 and 13.6 eV, relative to the background interstellar radiation field (ISRF; Habing 1968). Typically, UV dissociation is much smaller than dissociative recombination in an XDR. If we combine equation (2) with the dissociative recombination reaction rate, we can solve for the SH$^+$ density in steady state:

$$n(SH^+) = \frac{0.00165n(S)[7.9n(H_3^+)+n(HCO^+)]+\dfrac{k_1}{2\times10^{-7}}n(S^{+2})n(H_2)}{\left(\dfrac{T}{300}\right)^{-0.5}n(e)}. \qquad (3)$$

SH$^+$ leads to the formation of other sulfur-bearing molecules, including SH and H$_2$S$^+$. SH$^+$ is thought to be an excellent XDR diagnostic (Stäuber et al. 2005). Even without SH$^+$ formation via (1d), SH$^+$ is enhanced in XDRs due to increased ionization produced by X-rays. The key question we set out to answer is "can $S^{+2}$ + $H_2$ lead to further $SH^+$ enhancement in an XDR, and if so by how much?"

## 3 Computational Details

To quantify the effect of $S^{+2}$ chemistry on $SH^+$ formation, we performed a 1D, plane parallel geometry calculation using the spectral synthesis code Cloudy (Ferland et al. 1998). We briefly summarize the essential features of our model here, and point the reader to recent work (Abel et al. 2005, Shaw et al. 2005, van Hoof et al. 2004, and Röllig et al. 2007) and the Cloudy website[1] for a complete description of how Cloudy treats physical processes in XDRs and PhotoDissociation Regions (PDRs).

Our model is designed to be a simple calculation of an XDR surrounding the Narrow Line Region (NLR) of an AGN. Our calculations begin with all H in the form of $H^+$, and end when the fraction of H in $H_2$ exceeds 95%. The $2 \times n(H_2)/n(H_{total}) = 0.95$ point corresponds to $A_V = 10$ mag. We consider hydrogen densities $n_H = 10^3$, $10^4$, and $10^5$ cm$^{-3}$. We parameterize the AGN Spectral Energy Distribution (SED) using the SED given in Korista et al. (1997). This continuum is characterized by four parameters; the temperature ($T$) of the "Big Bump", the ratio of X-ray to UV flux ($\alpha_{ox}$), the low-energy slope of the Big Bump continuum ($\alpha_{uv}$), and the slope of the X-ray continuum ($\alpha_x$). For our calculations, we use $T = 10^6$ K, $\alpha_{ox} = 10^{-1.4}$, $\alpha_{uv} = 10^{-0.5}$, and $\alpha_x = 10^{-1.0}$. The intensity of the hydrogen ionizing radiation is defined in terms of the dimensionless ionization parameter $U = \dfrac{Q(H)}{4\pi R^2 n_H c}$, which we set to $10^{-2}$. The choice of $n_H$ and $U$ is meant to represent typical parameters for an AGN. We consider graphite and silicate dust grains, with an MRN grain size distribution (Mathis, Rumpl, & Nordsieck 1977). Opacities are taken from Martin & Rouleau (1991). Our choice of grain sizes and opacities yields a ratio of total to selective extinction typical of the Galactic ISM, $R_V = 3.1$. We scale the grain abundance such that $A_V/N_H = 5 \times 10^{-22}$ mag cm$^2$, also typical of the local ISM (Bohlin, Savage, & Drake 1978). Our assumed gas-phase abundances relative to hydrogen are an average of the local ISM from the work of Cowie & Songaila (1986) and Savage & Sembach (1996). For the most important element in this work, sulfur, we use S/H = $3.3 \times 10^{-5}$. We include the effects of cosmic rays, with an assumed cosmic ray ionization rate $\zeta_{cr} = 5 \times 10^{-17}$ s$^{-1}$.

---

[1] www.nublado.org

Given these parameters, the variable in our model is the branching ratio of the products in (1), with the total rate coefficient kept fixed at $1.6 \times 10^{-9}$ cm$^3$ s$^{-1}$. We allowed for two possible channels: formation of SH$^+$ via (1d) and single charge transfer, the sum of (1a) and (1b). The actual products in (1a) – (1c) have no effect on our results. We then vary the branching ratio between the two channels by multiplying this rate by coefficients $f_1$ and $f_2$, such that:

$$\text{Rate}[(1d)] = f_1 \times 1.6 \times 10^{-9} \text{ cm}^3 \text{ s}^{-1} \quad (4)$$

$$\text{Rate}[(1a)+(1b)] = f_2 \times 1.6 \times 10^{-9} \text{ cm}^3 \text{ s}^{-1} \tag{5}$$

with $f_1 + f_2 = 1$. We allow $f_1$ to vary from 1 (all $S^{+2} + H_2$ reactions lead to $SH^+$) to $10^{-9}$, in increments of 1 dex. We assume that the rate coefficient and branching ratios for process (1) do not depend on temperature. This is reasonable for a doubly-charged ion system, but in the worst case implies an upper limit that could decrease by a factor of ~2 at the lowest considered temperatures. In addition to reaction (1), we include many other $SH^+$ formation/destruction processes, the important reactions are summarized in Section 2.

## 4 Results & Discussion

Figure 1 shows the predicted $SH^+$ column density as a function of the branching ratio and $n_H$. The total production rates of the reactions used to derive equations (2) and (3), along with the radiative association reaction $S^+ + H \rightarrow SH^+ + h\nu$ (Stancil et al. 2000), which is not included in our Cloudy model, are displayed as a function of depth in Figure 2. From our results, it is clear $SH^+$ formation via (1d) may be a dominant channel in XDRs. Without $S^{+2}$ chemistry, our model predicts $\log[N(SH^+)] = 10^{11.6}$ cm$^{-2}$. $S^{+2}$ chemistry is unimportant to $SH^+$ formation until the branching ratio reaches ~1%. Once the branching ratio exceeds 1%, there is a sharp increase in $N(SH^+)$, with an increase of 1 dex for $f_1 = 0.1$ and 2 dex if $f_1 = 1$. This increase is somewhat smaller (but still significant) for $n_H = 10^5$ cm$^{-3}$ with increases of about 2.5 and 15 for $f_1 = 0.1$ and 1, respectively.

Figure 2 illustrates reaction (1d) is the dominant SH$^+$ formation process at all depths, when $f_1$ = 1.  Even when the branching ratio is 1%, the formation rate of reaction (1d) equals the formation rate via $H_3^+$, which explains the factor of 2 increase in $N$(SH$^+$) shown for a 1% branching ratio in Figure 1.  As long as the rate of SH$^+$ formation via S$^{+2}$ exceeds ~2×10$^{-11}$ cm$^3$ s$^{-1}$, our calculations predict this to be the dominant channel leading to SH$^+$.  Examination of the interaction potentials, deduced empirically from asymptotic energies for all channels in reaction (1) including excited states, suggests that the probabilities for each product channel, (1a) – (1d), are roughly equal.  An $f_1$ of ~0.25 is therefore plausible, but detailed calculations are needed to confirm this estimate.  Since this branching ratio leads to a more than 1 dex increase in the SH$^+$ abundance, it is very likely that doubly ionized chemistry is critical in the formation of SH$^+$ in XDRs.

Figure 3 shows the density of H$^+$, H$^0$, H$_2$, SH$^+$, and S$^{+2}$ as a function of $A_V$.  We show two scenarios, a calculation using the CGK rate (that was also used to produce Figures 1 and 2), and a calculation where S$^{+2}$ + H$_2$ is not considered.  We also plot the gas temperature as a function of $A_V$.  Our calculation shows the broad H/H$_2$ transition region characteristic of XDR models (Meijerink & Spaans 2005).  However, what is most interesting is the difference in the S$^{+2}$ abundance with depth.  With the CGK rate coefficient, S$^{+2}$ destruction by H$_2$ dominates over electron recombination once the gas becomes molecular.  Not including this process, or using the value of $10^{-15}$ cm$^3$ s$^{-1}$ listed in Y97, allows the S$^{+2}$ abundance to remain high out to $A_V$ = 10 mag. The total S$^{+2}$ column density predicted with CGK is log[$N$(S$^{+2}$)] = 16.91, or ~0.5 dex lower than $N$(S$^{+2}$) predicted when the Y97 rate coefficient is used, log[$N$(S$^{+2}$)] = 17.43. In both scenarios, over half of the total column density comes from regions where the H$^+$/H$_{total}$ fraction is less than 1%. With the CGK value for process (1), the XDR contributes about 50% of the S$^{+2}$ column density, while when CGK is not used the XDR contribution is closer to 90%. Overall, Figure 3 shows, regardless of the fraction of S$^{+2}$ + H$_2$ reactions that form SH$^+$, the CGK rate is very important in determining the ionization structure of sulfur in an XDR and should be included in any XDR calculation.

The possibility of SH$^+$ formation through S$^{+2}$ is intriguing, with important consequences for our understanding of the chemistry of a wide variety of environments. Since SH$^+$ leads to the formation of other molecules such as H$_2$S, there could be an entire series of sulfur-bearing molecules that are initiated by S$^{+2}$, instead of S or S$^+$. We determined what the branching ratio must be (>1%) so that S$^{+2}$ is the dominant formation pathway for SH$^+$ and other sulfur-bearing molecules in an XDR. An experimental or theoretical determination of the branching ratio is therefore needed. Observationally, Herschel should be able to detect SH$^+$ in the atomic/molecular gas near an AGN. If Herschel cannot detect SH$^+$, then it is likely that SH$^+$ does not form through S$^{+2}$ chemistry. We should be able to observe SH$^+$ in emission in galaxies dominated by AGN activity. Since SH$^+$ chemistry is initiated by X-rays, searches for SH$^+$ in UltraLuminous Infrared

Galaxies (ULIRGs) may be another way to detect embedded AGNs in ULIRGs. While Magnani & Salzer (1989, 1991) could only obtain upper limits on $SH^+$ in the diffuse ISM, this work and the work of Stäuber et al. (2005) demonstrate that the likely places to look for $SH^+$ in diffuse gas are near regions where the X-ray background is enhanced.

If the branching ratio turns out to exceed 1%, and $S^{+2}$ is important to forming $SH^+$, then it is likely other doubly ionized species could lead to molecule formation. Do ions like $O^{+2}$, $C^{+2}$, $Si^{+2}$, $N^{+2}$, and $Cl^{+2}$ lead to efficient formation of $OH^+$, $CH^+$, $SiH^+$, $NH^+$, and $HCl^+$? Each of these molecules leads to the formation of a wide variety of other molecules. Clearly, investigations into $X^{+2} + H_2 \rightarrow XH^+ + H^+$ reactions should be made, and the results implemented into XDR codes.

The presence of $S^{+2}$ in the $H^0$ region, but not the $H_2$ region, may be very important to observations of AGNs and ULIRGs. One of the most important diagnostics in the Spitzer wavelength window is the [S III] 18.7 / 33.5 μm intensity ratio. The diagnostic power of the [S III] ratio hinges on the fact that the temperature of the gas is much higher than the excitation temperature of each line, and that the level populations are determined by collisions. Our calculations find that about 50% of $N(S^{+2})$ is co-spatial with $H^0$. This means there could be significant 18.7 and 33.5 μm emission that does not come from the ionized gas but instead from the XDR, and this emission would be due to $H^0$ collisions and not electrons. Figure 3 shows the gas temperature in the $H^0$ region is a few thousand K, which is still higher than the excitation temperature of either line and allows for efficient [S III] emission. Therefore, the XDR component to [S III] could potentially diminish the effectiveness of the [S III] ratio as a density diagnostic in AGN or ULIRGs with an embedded AGN. An XDR component to [S III] emission could potentially explain some of the observations of Dudik et al. (2007), who found [S III] ratios below the low-density limit of 0.45 in 13 of 33 galaxies, although Dudik et al. (2007) suggested that aperture effects could be a plausible explanation.

In addition to the [S III] density ratio, the [S IV] 10.5 μm/[S III] 18.7, 33.5 μm diagnostic would be affected by an XDR component. A combined H II region + XDR model would predict a lower [S IV]/[S III] ratio than a pure H II region calculation. No laboratory data exist on the collisional excitation rates of [S III] via $H^0$. Our results point to a clear need to determine these rates in order to model the observed Spitzer and SOFIA spectrum of AGNs and ULIRGs accurately.

## 5 Conclusions

We studied the potential effects of $S^{+2} + H_2$ on the formation of $SH^+$ and destruction of $S^{+2}$ in XDRs. The important results of this work include:

- $S^{+2} + H_2 \rightarrow SH^+ + H^+$ could play a major role in the formation of $SH^+$ and other sulfur-bearing molecules in X-ray irradiated gas. If the percentage of

- $S^{+2} + H_2$ reactions that lead to $SH^+$ formation exceeds 1%, then this process becomes the dominant reaction leading to $SH^+$. Laboratory measurements of this branching ratio, combined with Herschel observations of AGNs and ULIRGs, should help to determine the importance of $S^{+2}$ chemistry in XDRs.

- The total CGK rate coefficient for $S^{+2}$ by $H_2$ should be used in any XDR calculation, with the consequence that $S^{+2}$ ions will not be significant in the $H_2$ region, but will be in the $H^0$ region. The presence of $S^{+2}$ in the $H^0$ region of an XDR may affect the interpretation of Spitzer observations of the [S III] density diagnostic and [S IV]/[S III] ratios, which are often assumed to be $H^+$ region diagnostics. Theoretical investigations into the magnitude of this effect are currently inhibited by the uncertainty in the collisional excitation of $S^{+2}$ by $H^0$. Determining the collisional rates and including them in XDR models is therefore important in interpreting Spitzer observations of AGNs and potentially ULIRGs.

- If $SH^+$ formation via $S^{+2}$ turns out to be viable, then the rate of other molecules such as $CH^+$, $OH^+$, $NH^+$, $SiH^+$, and $HCl^+$ via doubly ionized chemistry should be explored with XDR models in conjunction with the needed laboratory studies.

Acknowledgements: This material is based upon work supported by the National Science Foundation under Grant No. 0094050, 0607497 to The

University of Cincinnati. NPA also acknowledges the University of Cincinnati for computational support, in addition to the University of Kentucky and Miami of Ohio University for a generous allotment of time on their respective supercomputing clusters. PCS acknowledges support from NASA grant NNG06J11G. We would like to thank the anonymous referee for a careful reading of the manuscript.

# 7 Figures

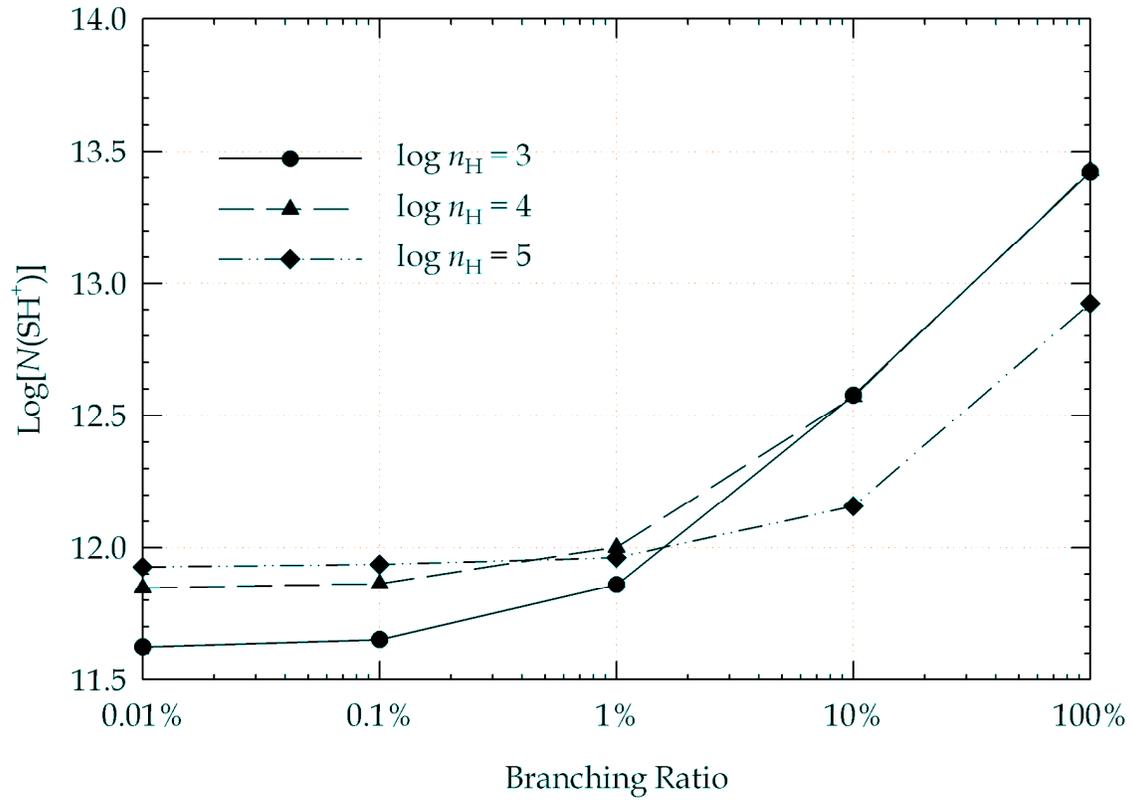

Figure 1 Variation in $N(SH^+)$ as a function of the percentage of $S^{+2}$ + $H_2$ reactions that form $SH^+$ and hydrogen density ($n_H$).

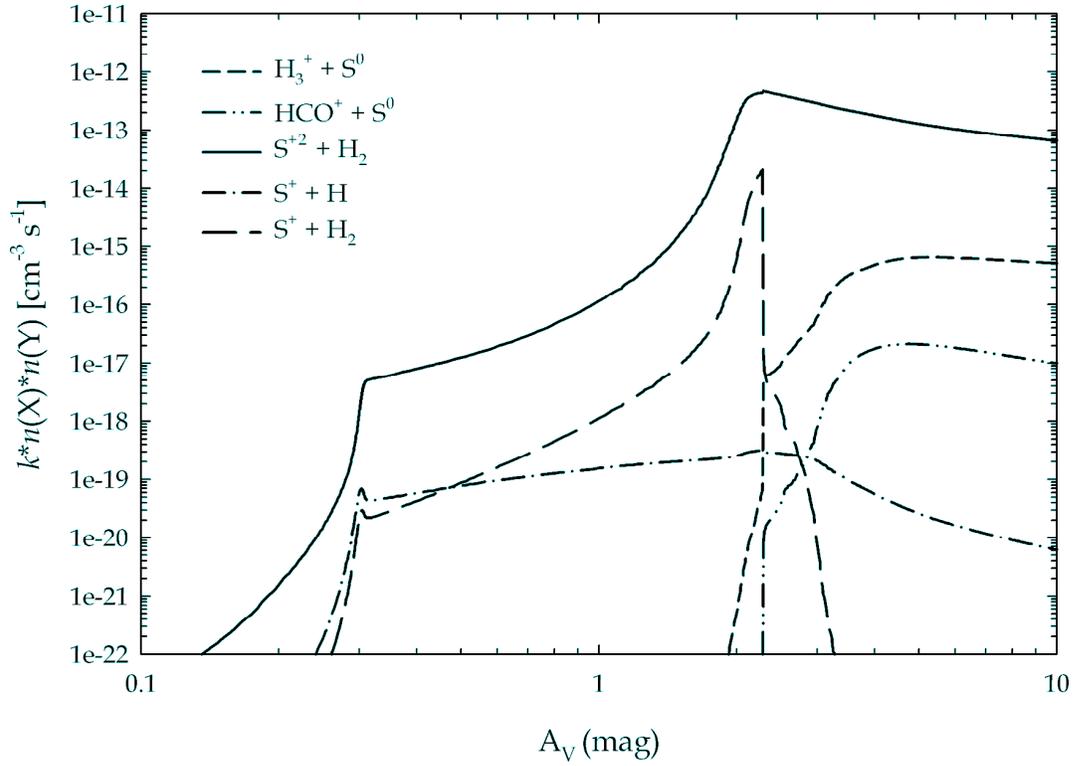

Figure 2 Important SH⁺ formation rates, for the case where 100% of $S^{+2} + H_2$ reactions lead to SH⁺ formation and for $n_H = 10^3$ cm⁻³. To compare the $S^{+2} + H_2$ formation rate to the other rates for different branching ratios, the rate shown here should be scaled by the branching ratio.

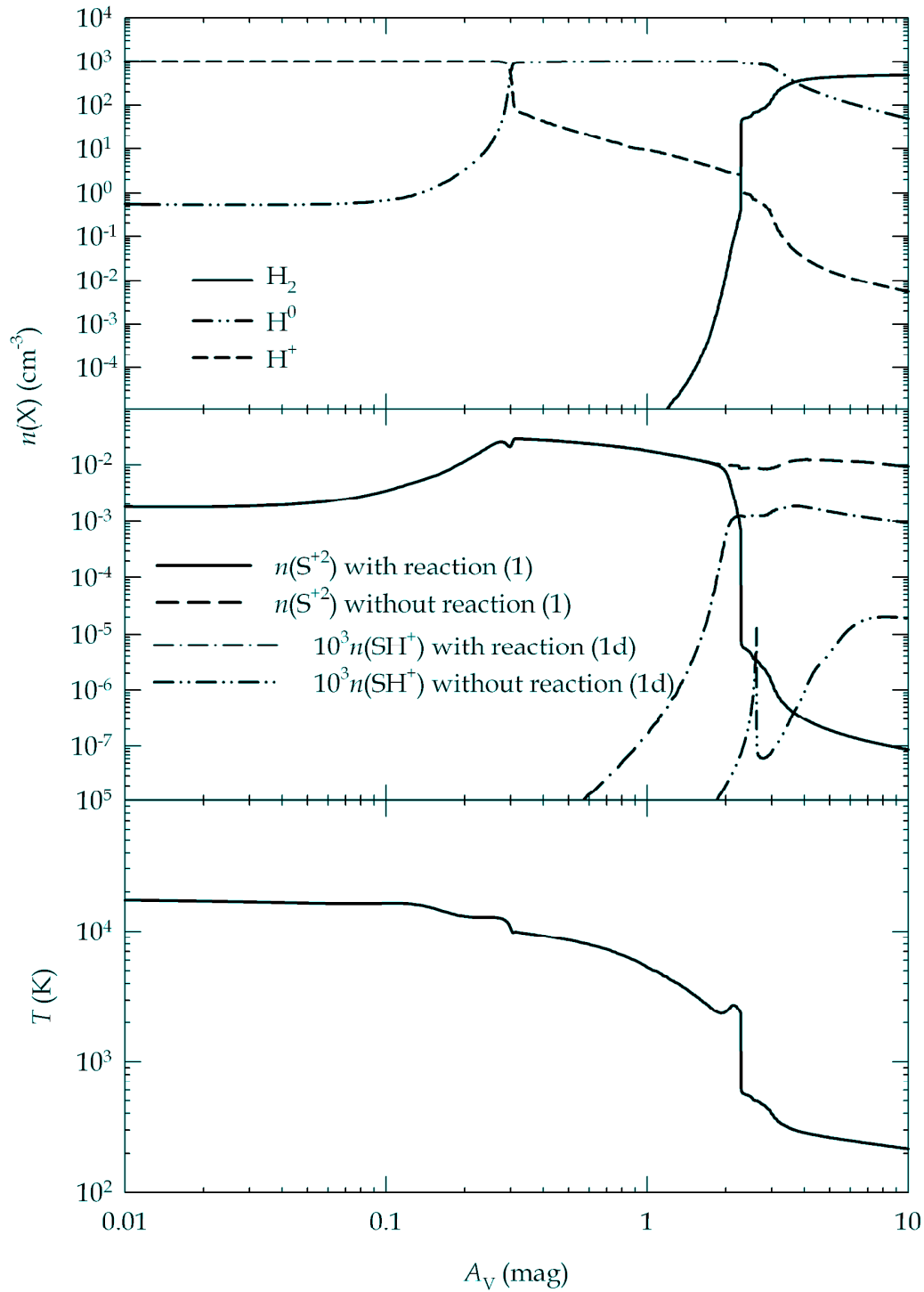

Figure 3 $S^{+2}$ and $SH^+$ density as a function of $A_V$, with and without the CGK rate coefficients, along with the $H^+$, $H^0$, $H_2$ densities and the gas temperature.